\documentclass[preprint,preprintnumbers,amsmath,amssymb]{revtex4}
\usepackage{graphicx}
\usepackage{dcolumn}
\usepackage{bm}
\usepackage{rotating}
\begin{document}
\newcommand{\n}{\noindent}
\newcommand{\e}{\mbox{e}}
\date{\today}

\title[Nonlocal connections]
{A nonlocal connection between certain linear and nonlinear
ordinary differential equations : Extension to coupled equations}

\author{  R. Gladwin Pradeep} \author{V. K. Chandrasekar}
\author{M. Senthilvelan}
\author{M. Lakshmanan}
\email{lakshman@cnld.bdu.ac.in}
\affiliation{Centre for Nonlinear Dynamics, Department of Physics,
Bharathidasan University, Tiruchirappalli - 620 024, India }

\begin{abstract}
Identifying integrable coupled nonlinear ordinary differential equations (ODEs) of dissipative type and deducing their general solutions are some of the challenging tasks in nonlinear dynamics.  In this paper we undertake these problems and unearth two classes of integrable coupled nonlinear ODEs of arbitrary order.  To achieve these goals we introduce suitable nonlocal transformations in certain linear ODEs and generate the coupled nonlinear ODEs.  In particular, we show that the problem of solving these classes of coupled nonlinear ODEs of any order, effectively reduces to solving a single first order nonlinear ODE.  We then describe a procedure to derive explicit general solutions for the identified integrable coupled ODEs, when the above mentioned first order nonlinear ODE reduces to a Bernoulli equation.  The equations which we generate and solve include the two coupled versions of modified Emden equations (in second order), coupled versions of Chazy equations (in third order) and their variants, higher dimensional coupled Ricatti and Abel chains as well as a new integrable chain and higher order equations.
\end{abstract}
\pacs{02.30.Hq, 02.30.Ik, 05.45.-a}

\maketitle
\section{Introduction}

During the past few years considerable interest has been shown towards
identifying a differential sequence of ordinary differential equations (ODEs)
and studying their integrability and symmetry properties{\footnotesize$^{1,2}$}.  An integrable differential sequence is a set of ODEs of which each equation in the sequence is integrable.  Here by an integrable $n^{\mbox{th}}$ order ODE we mean that an equation for which the general solution $x=f(t,c_1,\ldots,c_n)$, where $c_i$, $i=1,2,\ldots,n$, are arbitrary constants, can be explicitly found (to within quadratures).

To identify such a sequence of equations a general algorithm has been proposed by three of the present authors with Kundu{\footnotesize$^{3}$}.  The algorithm not only yields the well known Ricatti and Abel chains but also produces chains{\footnotesize$^{4,5}$} which involve inverse polynomials and more general systems.  On the other hand, several recent studies have been confined to generation and analysis of only Ricatti and Abel chains{\footnotesize$^{2}$}.

The Ricatti chain is defined by $\mathbb{D}_R^mx=0$, $m=0,1,2,3,...$, where $\mathbb{D}_R=\frac{d}{dt}+kx$, $k\in\mathbb{R}$ so that the action of $\mathbb{D}_R$ generates the following family of differential equations{\footnotesize$^{2}$}, namely
\begin{eqnarray}
&&\mathbb{D}_R^0x=x,\nonumber\\
&&\mathbb{D}_Rx=\left(\frac{d}{dt}+kx\right)x=\dot{x}+kx^2,\nonumber\\
&&\mathbb{D}_R^2x=\left(\frac{d}{dt}+kx\right)^2x=\ddot{x}+3kx\dot{x}+k^2x^3,\\
 &&\mathbb{D}_R^3x=\left(\frac{d}{dt}+kx\right)^3x=
\dddot{x}+4kx\ddot{x}+6k^2x^2\dot{x}+3k\dot{x}^2+k^3x^4,\nonumber
\end{eqnarray}
and so on.  The second member in this family is the Ricatti equation and the third member is the modified Emden equation and the fourth member is one of the subcases of the Chazy equation and so on.

On the other hand considering the differential operator in the form
$\mathbb{D}_A=\frac{d}{dt}+kx^2$, $k\in\mathbb{R}$, one can generate  a hierarchy of higher order Abel equations{\footnotesize$^{2}$}, namely

\begin{eqnarray}
&& \mathbb{D}_A^0x=x,\nonumber\\
&& \mathbb{D}_Ax=\left(\frac{d}{dt}+kx^2\right)x=\dot{x}+kx^3,\nonumber\\
&&\mathbb{D}_A^2x=\left(\frac{d}{dt}+kx^2\right)^2x=\ddot{x}+4kx^2\dot{x}+k^2x^5,\\
&& \mathbb{D}_A^3x=\left(\frac{d}{dt}+kx^2\right)^3x=
\dddot{x}+5kx^2\ddot{x}+8kx\dot{x}^2+9k^2x^4\dot{x}+k^3x^7,\nonumber
\end{eqnarray}
etc.  The second member is a special case of the Bernoulli equation and the third member is a generalized van-der Pol oscillator equation and the fourth member is a subcase of the Chazy equation.  It has been shown that all the equations in each of the hierarchy posses certain common properties that make them interesting both from physical and mathematical points of view.  For example, as far as the Ricatti chain is concerned (i) all the equations in this chain admit a maximal number of Lie point symmetries, (ii) every equation of order $m$ in this hierarchy can be linearized and transformed into a linear ODE of order $m+1$ and (iii) the dimensional reduction of a linear equation of order $m+1$ leads to the Ricatti equation of order $m$.  As far as the Abel chain is concerned all the equations in this hierarchy posses a family of first integrals $J_{tk}$ that depend on time as a polynomial.  In addition to the above, specific equations in both the categories have also been investigated in detail.  In particular, the second equation in the Ricatti chain, namely the modified Emden equation, has been a central attraction for mathematicians and physicists for more than a century (for more details see Refs.[6-10]).

As we mentioned earlier, in Ref. [3] it has been demonstrated that one can generate both the Ricatti and Abel chains of equations from the damped linear harmonic oscillator equation, $\ddot{U}+c_1\dot{U}+c_2U=0$, where $c_1$ and $c_2$ are arbitrary parameters, and higher order linear ODEs by introducing a nonlocal transformation, $U=x^{\alpha}e^{\int(\beta(t)x^m+\gamma(t))}dt$, where $\alpha$ and $m$ are constants and $\beta$ and $\gamma$ are arbitrary functions of $t$.  Substituting the above nonlocal transformation in the damped harmonic oscillator equation and restricting the parameters appropriately ($\alpha=1,\,\beta=k,\,\gamma=0,\,c_1,\,c_2=0,\,m=1$) one can get the third member in the Ricatti chain.  Similarly, taking $\alpha=1,\,\beta=k,\,\gamma=0,\,c_1,\,c_2=0,\,m=2$ one can get the third member in the Abel chain.  Now considering a third order linear ODE $\dddot{U}+c_1\ddot{U}+c_2\dot{U}+c_3U=0$, where $c_1,\,c_2,\,c_3$ are constants, and substituting the above nonlocal transformation in this equation, and restricting the parameters suitably, one can get the fourth member of both the chains.  The authors have also presented a straightforward method of finding explicit general solution for all the equations belonging to these chains.

Even though several works have been dedicated to explore the physical and mathematical properties of these two chains no attempt has been made to identify/generate a sequence of coupled integrable ODEs and construct their general solutions, which are in general difficult tasks.  The aim of this paper is to make some progress in these directions and present some of the important results regarding how to generate two different classes of integrable coupled ODEs and the method of constructing their general solutions.  We show that the problem of solving the chain of coupled integrable ODEs generated by this procedure can be ultimately reduced to solving a single first order nonlinear ODE.  In other words, the problem of constructing the solution of the system of nonlinear coupled ODEs effectively reduce to deducing the solution of the equivalent first order nonlinear ODE.  We deduce the solution of this first order equation for the parametric choice for which it reduces to an integrable Bernoulli equation.  We find that for specific choices of system parameters the system of coupled nonlinear ODEs exhibit isochronous oscillations, which is an interesting area of study in its own merit{\footnotesize$^{11}$}.  In the process of identifying two new integrable class of equations we have obtained a higher dimensional generalization of the Ricatti and Abel chains.  The two dimensional Ricatti chain has been identified as
\begin{eqnarray}
&&\mathbb{D}_{R}^0
\left(
\begin{array}{l}
x_1\\
x_2
\end{array}
\right)\,\,\Rightarrow
\begin{array}{l}
x_1=0,\\
x_2=0.
\end{array}\label{chaineq1}\\
&&\mathbb{D}_{R}^1
\left(
\begin{array}{l}
x_1\\
x_2
\end{array}
\right)
\,\,\Rightarrow
\begin{array}{l}\dot{x}_1+(a_1x_1+a_2x_2)x_1=0,\\
\dot{x}_2+(a_1x_1+a_2x_2)x_2=0.
\end{array}\\
&&\mathbb{D}_{R}^2
\left(
\begin{array}{l}
x_1\\
x_2
\end{array}
\right)\,\,\Rightarrow
\begin{array}{l}
\ddot{x}_1+2(a_1x_1+a_2x_2)\dot{x}_1+(a_1\dot{x}_1+a_2\dot{x}_2)x_1
+(a_1x_1+a_2x_2)^2x_1=0,\\
\ddot{x}_2+2(a_1x_1+a_2x_2)\dot{x}_2+(a_1\dot{x}_1+a_2\dot{x}_2)x_2
+(a_1x_1+a_2x_2)^2x_2=0.\\
\end{array}\label{cmee}\\
&&\mathbb{D}_{R}^3
\left(
\begin{array}{l}
x_1\\
x_2
\end{array}
\right)\Rightarrow
\begin{array}{l}
\dddot{x}_1+(3(a_1x_1+a_2x_2))\ddot{x}_1+x(a_1\ddot{x}_1+a_2\ddot{x}_2)
+3(a_1\dot{x}_1+a_2\dot{x}_2)\dot{x}_1\\
\qquad
+(a_1x_1+a_2x_2)(3\dot{x}_1(a_1x_1+a_2x_2)
+3x_1(a_1\dot{x}_1+a_2\dot{x}_2)+(a_1x_1+a_2x_2)^2x_1)=0,\\
\dddot{x}_2+(3(a_1x_1+a_2x_2))\ddot{x}_2
+y(a_1\ddot{x}_1+a_2\ddot{x}_2)
+3(a_1\dot{x}_1+a_2\dot{x}_2)\dot{x}_2
\\
\qquad
+(a_1x_1+a_2x_2)(3\dot{x}_2(a_1x_1+a_2x_2)
+3x_2(a_1\dot{x}_1+a_2\dot{x}_2)+(a_1x_1+a_2x_2)^2x_2)=0,
\end{array}\label{chazy}
\end{eqnarray}
and so on, where
$\mathbb{D}_{R}^l=\left(\frac{d}{dt}+(a_1x_1+a_2x_2)\right)^l.$
This two dimensional Ricatti chain can be further generalized to arbitrary dimensions and the details are given in Sec. \ref{section7}.  
The two dimensional Abel chain is given as
\begin{eqnarray}
&&\hspace{-0.7cm}\mathbb{D}_{A}^0
\left(
\begin{array}{l}
x_1\\
x_2
\end{array}
\right)\,\,\Rightarrow
\begin{array}{l}
x_1=0,\\
x_2=0.\label{abelchain1}
\end{array}\\
&&\hspace{-0.7cm}\mathbb{D}_{A}^1
\left(
\begin{array}{l}
x_1\\
x_2
\end{array}
\right)\,\,\Rightarrow
\begin{array}{l}\dot{x}_1+(a_1x_1^2+a_2x_2^2)x_1=0,\\
\dot{x}_2+(a_1x_1^2+a_2x_2^2)x_2=0.
\end{array}\\
&&\hspace{-0.7cm}\mathbb{D}_{A}^2
\left(
\begin{array}{l}
x_1\\
x_2
\end{array}
\right)\,\,\Rightarrow
\begin{array}{l}
\ddot{x}_1+(2(a_1x_1^2+a_2x_2^2))\dot{x}_1
+x_1((a_1x_1^2+a_2x_2^2)^2+2a_1x_1\dot{x}_1+2a_2x_2\dot{x}_2)=0,\\
\ddot{x}_2+(2(a_1x_1^2+a_2x_2^2))\dot{x}_2
+x_2((a_1x_1^2+a_2x_2^2)^2+2a_1x_1\dot{x}_1+2a_2x_2\dot{x}_2)=0.
\end{array}\label{dvp}\\
&&\hspace{-0.7cm}\mathbb{D}_{A}^3
\left(
\begin{array}{l}
x_1\\
x_2
\end{array}
\right)\,\,\Rightarrow
\begin{array}{l}
\dddot{x}_1+3(a_1x_1^2+a_2x_2^2)\ddot{x}_1+6(a_1x_1\dot{x}_1+a_2x_2\dot{x}_2)(\dot{x}_1+x_1(a_1x_1^2+a_2x_2^2))\\
+2x_1(a_1\ddot{x}_1+a_2x_2\ddot{x}_2)+2a_1x_1\dot{x}_1^2+3(a_1x_1^2+a_2x_2^2)^2+x_1(a_1x_1^2+a_2x_2^2)^3=0,\\
\dddot{x}_2+3(a_1x_1^2+a_2x_2^2)\ddot{x}_2+6(a_1x_1\dot{x}_1+a_2x_2\dot{x}_2)(\dot{x}_2+x_2(a_1x_1^2+a_2x_2^2))\\
2x_2(a_1\ddot{x}_1+a_2x_2\ddot{x}_2)+2a_2x_2\dot{x}_2^2+3(a_1x_1^2+a_2x_2^2)^2+x_2(a_1x_1^2+a_2x_2^2)^3=0,
\end{array}\label{abelchain4}
\end{eqnarray}
and so on, where
$\mathbb{D}_{A}^l=\left(\frac{d}{dt}+(a_1x_1^2+a_2x_2^2)\right)^l$.  
Some of the interesting equations in the above chains are the coupled modified Emden equations (\ref{cmee}), coupled generalization of Chazy type equation (\ref{chazy}) and the coupled generalized Duffing-van der Pol oscillator equations (\ref{dvp}), the details of which are given in Secs. \ref{section3} and \ref{section6}.  The $n$-dimensional versions of the above two chains are in Sec. \ref{section7}.

In addition to the $n$-dimensional Ricatti and Abel chains, we have also identified a new integrable chain.  The simplified two dimensional version of this new integrable chain is given as
\begin{eqnarray}
&&\mathbb{D}_N^0 
\left(
\begin{array}{l}
x_1\\
x_2
\end{array}
\right)\Rightarrow
\begin{array}{l}
x_1=0,\\
x_2=0.
\end{array}\label{newchain1}\\
&&\mathbb{D}_N^1
\left(
\begin{array}{l}
x_1\\
x_2
\end{array}
\right)\Rightarrow
\begin{array}{l}
\dot{x}_1+a_1x_1^2x_2=0,\\
\dot{x}_2+a_1x_1x_2^2=0.
\end{array}\\
&&\mathbb{D}_N^2
\left(
\begin{array}{l}
x_1\\
x_2
\end{array}
\right)\Rightarrow
\begin{array}{l}
\ddot{x}_1+2a_1x_1x_2\dot{x}_1+a_1x_1^3x_2^2+a_1(\dot{x}_1x_2+x_1\dot{x}_2)=0,\\
\ddot{x}_2+2a_1x_1x_2\dot{x}_2+a_1x_1^2x_2^3+a_1(\dot{x}_1x_2+x_1\dot{x}_2)=0.
\end{array}\label{newchain3}\\
&&\mathbb{D}_N^3
\left(
\begin{array}{l}
x_1\\
x_2
\end{array}
\right)\Rightarrow
\begin{array}{l}
\dddot{x}_1+2a_1(x_1x_2\ddot{x}_1+x_1\dot{x}_1\dot{x}_2+\dot{x}_1^2x_2)
+a_1(3x_1^2x_2\dot{x}_1+x_1^3\dot{x}_2)\\
\qquad\qquad\quad+a_1(\ddot{x}_1x_2+x_1\ddot{x}_2+2\dot{x}_1\dot{x}_2)=0,\\
\dddot{x}_2+2a_1(x_1x_2\ddot{x}_2+x_2\dot{x}_1\dot{x}_2+\dot{x}_1x_2^2)
+a_1(3x_1x_2^2\dot{x}_2+x_1\dot{x}_2^3)\\
\qquad\qquad\quad+a_1(\ddot{x}_1x_2+x_1\ddot{x}_2+2\dot{x}_1\dot{x}_2)=0,
\end{array}\label{newchain4}
\end{eqnarray}
and so on, where $\mathbb{D}_N^{(l)}=\left(\frac{d}{dt}+a_1x_1x_2\right)^l$. The above integrable chain can be further generalized to $n$-dimensions and the details are also given in Sec. \ref{section7}.  

The plan of the paper is as follows.  In Sec. \ref{section2}, we point out the nonlocal connection between two uncoupled damped linear harmonic oscillators and certain coupled nonlinear oscillator equations.  We also present a method of identifying suitable nonlocal transformations for this purpose.  We then discuss the method of constructing the general solution for the identified set of nonlinear two coupled ODEs in Sec. \ref{section3}.  Generalizing this analysis, in Sec. \ref{section5}, we consider a system of $n$ uncoupled damped harmonic oscillator equations and generate a class of $n$ coupled nonlinear second order ODEs.  In Sec. \ref{section6}, we extend the procedure to identify a class of integrable third order coupled nonlinear ODEs.  To make our studies systematic here also we first consider a set of two coupled third order linear equations and by introducing the same nonlocal transformations we identify a class two coupled third order nonlinear ODEs.  This procedure is then extended to $n$-coupled third order ODEs.  In Sec. \ref{section7}, we show that the results obtained for second order and third order coupled nonlinear ODEs can be extended to arbitrary
$l^{\mbox{th}}$ order coupled nonlinear ODEs.  As special cases,  in addition to the $n$-dimensional generalization of Ricatti and Abel chains, we identify a new $n$-dimensional integrable chain of nonlinear ODEs.  In Sec. \ref{inhomogeneous} we consider a system of uncoupled inhomogeneous linear equations with variable coefficients and briefly discuss the applicability of our procedure. We summarize our results in Sec. \ref{section8}.
\section{Nonlocal connection between second order linear ODEs and coupled nonlinear ODEs}
\label{section2}
While studing the dynamics of a set of two coupled modified Emden equations {\footnotesize$^{12}$},
\begin{eqnarray}
&&\ddot{x}+2(k_1x+k_2y)\dot{x}+(k_1\dot{x}+k_2\dot{y})x+
(k_1x+k_2y)^2x+\lambda_1 x=0,\nonumber\\
&&\ddot{y}+2(k_1x+k_2y)\dot{y}+(k_1\dot{x}+k_2\dot{y})y+
(k_1x+k_2y)^2y+\lambda_2 y=0,\label{secondordercmee}
\end{eqnarray}
we have identified the fact that the system (\ref{secondordercmee})
can be transformed into two uncoupled harmonic oscillator equations, namely $\ddot{U}+\lambda_1U=0$ and $\ddot{V}+\lambda_2V=0$, where $\lambda_1,\lambda_2$ are constants,
through the nonlocal transformation $U=xe^{\int (k_1x+k_2y)dt}$ and $V=ye^{\int (k_1x+k_2y)dt}$.  Interestingly the solution for Eq. (\ref{secondordercmee}) can be derived from the known solution of the uncoupled linear equations $\ddot{U}+\lambda_1U=0$ and $\ddot{V}+\lambda_2V=0$, say $U=a(t)$ and $V=b(t)$.  This can be done by noting the fact that $x$ and $y$ are also related with $U$ and $V$ through the set of first order coupled nonlinear ODEs of the form
\begin{eqnarray}
\dot{x}=\frac{\dot{U}}{U}x-(k_1x+k_2y)x,\,\,\dot{y}=\frac{\dot{V}}{V}y-(k_1x+k_2y)y.\label{bern-mee}
\end{eqnarray}
Since $\frac{\dot{U}}{U}=\frac{\dot{a}}{a}$ and $\frac{\dot{V}}{V}=\frac{\dot{b}}{b}$ are known functions, Eqs. (\ref{bern-mee}) can be integrated straightforwardly and rearranging the constants suitably one can deduce the general solution for equation (\ref{secondordercmee}) in the form
\begin{eqnarray}
&&\qquad x(t)=\frac{A\sin(\omega_1 t+\delta_1)}
{1-\frac{Ak_1}{\omega_1}\cos(\omega_1 t+\delta_1)
-\frac{Bk_2}{\omega_2}\cos(\omega_2 t+\delta_2)},\omega_1=\sqrt{\lambda_1},\,
\omega_2=\sqrt{\lambda_2}\nonumber\\
&&\qquad y(t)=\frac{B\sin(\omega_2 t+\delta_2)}
{1-\frac{Ak_1}{\omega_1}\cos(\omega_1 t+\delta_1)
-\frac{Bk_2}{\omega_2}\cos(\omega_2 t+\delta_2)},\,\,\quad
\bigg|\frac{Ak_1}{\omega_1}+\frac{Bk_2}{\omega_2}\bigg|<1\label {cmee04b}
\end{eqnarray}
where $A,B,\delta_1,\delta_2$ are four arbitrary constants.
\subsection{Generalization}
\label{secnonlocal}
Now the question naturally arises as to whether one can generalize the nonlocal transformations and identify a chain of coupled nonlinear ODEs as in the case of the scalar case{\footnotesize$^{3,13}$}.  A systematic investigation reveals the fact that this can indeed be done.

For this purpose, we consider a set of two uncoupled damped linear harmonic oscillators, defined by the system of uncoupled second order linear ordinary differential equations (ODEs)
\begin{eqnarray}
\ddot{U}+c_{11}\dot{U}+c_{12}U=0,\,\,\ddot{V}+c_{21}\dot{V}+c_{22}V=0,
\label {slv01}
\end{eqnarray}
where $c_{11},\,c_{12},\,c_{21}$ and $c_{22}$ are arbitrary parameters for the present.  However, they can be as well functions of $t$ as we point out later in Sec. \ref{section8}.  Introducing a  nonlocal transformation,
\begin{eqnarray}
U=x^{\alpha} e^{\int f(x,y,t)dt},\qquad\,\, V=y^{\beta} e^{\int g(x,y,t)dt},\label{nonlocal-trans}
\end{eqnarray}
where $f$ and $g$ are two arbitrary functions of their arguments, in (\ref{slv01}) we obtain a set of two coupled second order nonlinear ODEs of the form
\begin{subequations}
\begin{eqnarray}
\addtocounter{equation}{-1}
\label{gen-non-coup}
\addtocounter{equation}{1}
&&\qquad \ddot{x}+(\alpha-1)\frac{\dot{x}^2}{x}+(2f+c_{11})\dot{x}
+\frac{x}{\alpha}\left[f^2+c_{11}f+c_{12}+\dot{f}\right]=0,\,\\
&&\qquad \ddot{y}+(\beta-1)\frac{\dot{y}^2}{y}+(2g+c_{21})\dot{y}+\frac{y}{\beta}\left[g^2+c_{21}g+c_{22}+\dot{g}\right]=0,
\end{eqnarray}
\end{subequations}
\noindent where $\displaystyle\dot{f}=\frac{df}{dt}=\frac{\partial f}{\partial t}+\dot{x}\frac{\partial f}{\partial x}+\dot{y}\frac{\partial f}{\partial y}$ and
$\displaystyle\dot{g}=\frac{dg}{dt}=\frac{\partial g}{\partial t}+\dot{x}\frac{\partial g}{\partial x}+\dot{y}\frac{\partial g}{\partial y}$.
\noindent Obviously choosing $\alpha=\beta=1$ and the functions $f$ and $g$ to be linear in $x$ and $y$, Eq. (\ref{gen-non-coup}) reduces to (\ref{secondordercmee}).

In order to find a family of integrable nonlinear
ODEs in the class (\ref{gen-non-coup}) which are connected with the
damped linear harmonic oscillator equations (\ref{slv01}) through the nonlocal transformation (\ref{nonlocal-trans})
and deduce the general solution for the coupled nonlinear ODEs from the solutions of the linear ODEs (\ref{slv01}), we find that the task essentially consists of two parts.  First one has to choose specific
forms of $f$ and $g$ and fix the target coupled nonlinear ODEs and then one can look for a suitable method to derive the general solution of the nonlinear ODEs.  In the
following we describe a procedure which successfully takes care of both the steps.

To start with we observe that the nonlocal transformation (\ref{nonlocal-trans}) yields a system
of two first order coupled nonlinear, nonautonomous ODEs of the form
\begin{eqnarray}
\dot{x}=\frac{x}{\alpha}\left[\frac{\dot{U}}{U}-f(x,y,t)\right],
\qquad\dot{y}=\frac{y}{\beta}\left[\frac{\dot{V}}{V}-g(x,y,t)\right],\label {generalriccati}
\end{eqnarray}
where $U$ and $V$ are the solutions of the linear ODEs (\ref{slv01}).  One may
note that since $\frac{\dot{U}}{U}$ and $\frac{\dot{V}}{V}$ are some known functions
of `$t$' the first terms on the right hand sides of both the equations have explicit time dependent factors.

Note that for the choice $f(x,y,t)=a_{11}(t)x+a_{12}(t)y$ and $g(x,y,t)=a_{21}(t)x+a_{22}(t)y$, Eq. (\ref{generalriccati}) reduces to the time dependent two dimensional Lotka-Volterra (LV) equation
\begin{eqnarray}
\dot{x}=x(a_{11}(t)x+a_{12}(t)y+b_1(t)),\,\,\dot{y}=y(a_{21}(t)x+a_{22}(t)y+b_2(t)),\label{lv}
\end{eqnarray}
where $b_1(t)=\frac{\dot{U}}{U}$ and $b_2(t)=\frac{\dot{V}}{V}$.
 The parameters $\alpha$, $\beta$ have been absorbed into $b_i$ and $a_{ij}$, $i,j=1,2$.  Studies have been carried out on the dynamics of this system{\footnotesize$^{14-16}$}.  However, we find that Eq. (\ref{lv}) is not integrable for arbitrary choice of the coefficients $a_{ij}$, and $b_i$, $i,j=1,2$, but for special forms they are integrable as we see below.

In general it is easier to integrate Eq. (\ref{generalriccati}) than to solve
Eq. (\ref{gen-non-coup}) and obtain $x(t)$ and $y(t)$.  The question now
boils down to the fact for what forms of $f$ and $g$ Eq. (\ref{generalriccati}) can be
integrated to yield the general solution.  While analysing the
form of Eq. (\ref{generalriccati}) we find that upon choosing $f$ and $g$ in a
symmetric form a first integral can be obtained by suitably rewriting and
integrating the resultant equations.  Of course the form of $f$ and $g$ have to be fixed in
such a way that the resultant set of coupled first order ODEs falls into a coupled Bernoulli family
of equations so that the general solution can be obtained.  To be specific, we identify two specific forms of $f$ and $g$.
  They are
\begin{eqnarray}
\FL(i)\quad f=(a_{1}(t)x^{p}+a_{2}(t)y^{q})^m+b_1(t),\quad g=(a_{1}(t)x^{p}+a_{2}(t)y^{q})^m+b_2(t)\label{form-f1}
\end{eqnarray}
with the parametric restriction $\frac{p\beta}{\alpha}=q$ (details of this parametric restriction are given in Sec. \ref{section3})
and
\begin{eqnarray}
\FL(ii)\quad f=(\sum_{i=1}^Na_i(t)x^{p_i}y^{q_i})^m+b_1(t),\quad g=(\sum_{i=1}^Na_i(t)x^{p_i}y^{q_i})^m+b_2(t)\label{form-g1},
\end{eqnarray}
with the parametric restrictions $\frac{p_i\beta}{\alpha}+q_i=k$, $i=1,2,\ldots,N$.  Here
$a_i$'s and $b_j$'s, $i=1,2,\ldots,N$, $j=1,2$, are arbitrary functions of $t$, and $p,\,q,\,p_i,\,q_i$ are real numbers satisfying the above parametric restrictions and $k$ is a constant.  For both the cases we find that the resultant Bernoulli equation can be integrated to yield the solution which in turn leads to the general solution of the nonlinear coupled second order ODEs with appropriate redefinition of the integration constants as we see in the following sections.

\section{Integrable Two coupled second order nonlinear ODEs}
\label{section3}
In this section we illustrate the procedure for two coupled second order ODEs.  The two different forms of $f$ and $g$  given by Eqs. (\ref{form-f1}) and (\ref{form-g1}) are treated separately as Case 1 and Case 2 in the following.

\noindent{\bf\underline{Case : 1}}

\noindent For the forms of $f$ and $g$ given in (\ref{form-f1}),  Eq. (\ref{generalriccati}) reduces to a system of coupled Bernoulli type equations,
\begin{subequations}
\begin{eqnarray}
\addtocounter{equation}{-1}
\label{coup-bernoulli-1}
\addtocounter{equation}{1}
\dot{x}=\frac{x}{\alpha}\left[\frac{\dot{U}}{U}-b_1(t)-(a_{1}(t)x^{p}+a_{2}(t)y^{q})^m\right],\\
\dot{y}=\frac{y}{\beta}\left[\frac{\dot{V}}{V}-b_2(t)-(a_{1}(t)x^{p}+a_{2}(t)y^{q})^m\right].
\end{eqnarray}
\end{subequations}
Substituting this form of $f$ and $g$ into Eq. (\ref{gen-non-coup}) we obtain the corresponding set of coupled nonlinear second order ODEs as
\begin{subequations}
\begin{eqnarray}
\addtocounter{equation}{-1}
\label{bernoulli-derived}
\addtocounter{equation}{1}
&&\ddot{x}+(\alpha-1)\frac{\dot{x}^2}{x}+\left[2((a_1x^p+a_2y^q)^m+b_1)+c_{11}\right]\dot{x}+\frac{x}{\alpha}\bigg[((a_1x^p+a_2y^q)^{m}+b_1)^2 +c_{12}
\nonumber\\
&&\qquad\quad+c_{11}((a_1x^p+a_2y^q)^m+b_1)+m\left(a_1x^p+a_2y^q\right)^{m-1}\bigg\{a_1px^{p-1}\dot{x}+a_2qy^{q-1}\dot{y}\nonumber\\
&&\qquad\qquad\qquad+a_{1t}x^p+a_{2t}y^q\bigg\}+b_{1t}\bigg]=0,\\
&&\ddot{y}+(\beta-1)\frac{\dot{x}^2}{x}+\left[2((a_1x^p+a_2y^q)^m+b_2)+c_{21}\right]\dot{y}+\frac{y}{\beta}\bigg[((a_1x^p+a_2y^q)^{m}+b_2)^2 +c_{22}
\nonumber\\
&&\qquad\quad+c_{21}((a_1x^p+a_2y^q)^m+b_2)+m\left(a_1x^p+a_2y^q\right)^{m-1}\bigg\{a_1px^{p-1}\dot{x}+a_2qy^{q-1}\dot{y}\nonumber\\
&&\qquad\qquad\qquad+a_{1t}x^p+a_{2t}y^q\bigg\}+b_{2t}\bigg]=0.
\end{eqnarray}
\end{subequations}

Solution of Eq. (\ref{bernoulli-derived}) can be written down by solving the  coupled
 Bernoulli type equations (\ref{coup-bernoulli-1}) provided the parametric condition $\frac{p\beta}{\alpha}=q$ is satisfied .
For this purpose, we multiply the first equation of (\ref{coup-bernoulli-1}) by $\alpha y$ and the
second by $\beta x$ and subtract the latter from the former to obtain
\begin{eqnarray}
\alpha y\dot{x}-\beta x\dot{y}=\left(\frac{\dot{U}}{U}-\frac{\dot{V}}{V}+b_2(t)-b_1(t)\right)xy.
\label{coup-bern-1}
\end{eqnarray}
Note that the general form of the solutions of the linear ODEs (\ref{slv01}) can be written down as
\begin{eqnarray}
U(t)=I_1e^{m_1t}+I_2e^{m_2t},\quad V(t)=I_3e^{m_3t}+I_4e^{m_4t},\label{2linsol}
\end{eqnarray}
where $I_1,\,I_2,\,I_3,$ and $I_4$ are four arbitrary constants, $m_1$ and $m_2$ are the solutions of the auxiliary equation $m^2+c_{11}m+c_{12}=0$, while $m_3$ and $m_4$ are the solutions of $m^2+c_{21}m+c_{22}=0$.
Now dividing both sides of (\ref{coup-bern-1}) by $xy$ and integrating the resultant equation (\ref{coup-bern-1}) one gets
\begin{eqnarray}
x(t)=\left(\frac{U}{V}y^{\beta}e^{\int(b_2-b_1)dt}\right)^{\frac{1}{\alpha}}.\label{xyrelation}
\end{eqnarray}
We also observe that the integration constant has been absorbed with the arbitrary constants in $\frac{U}{V}$.
Substituting (\ref{xyrelation}) in the second equation of Eq. (\ref{coup-bernoulli-1}) we get
\begin{eqnarray}
\dot{y}=\left(\frac{\dot{V}}{V}-b_2\right)\frac{y}{\beta}-
\frac{y}{\beta}\left[a_1\left(\frac{U}{V}e^{\int(b_2-b_1)dt}\right)^{\frac{p}{\alpha}}y^{\frac{p\beta}{\alpha}}
+a_2y^q\right]^m.\label{bern2}
\end{eqnarray}
The above equation reduces to the Bernoulli type equation
\begin{eqnarray}
\dot{y}=f(t)y-g(t)y^{mq+1}\label{bernoulli-type}
\end{eqnarray}
 for the choice
$\frac{p\beta}{\alpha}=q$.  Note that for the simple choice $\alpha=\beta$, we have the relation $p=q$.
Integrating now equation (\ref{bern2}) we obtain $y(t)$ and once $y(t)$ is known, $x(t)$ can be
obtained straightforwardly through (\ref{xyrelation}) (where again the integration constant has been absorbed into that of $U$ and $V$),
\begin{subequations}
\addtocounter{equation}{-1}
\label{gen-sol1-gen}
\addtocounter{equation}{1}
\begin{eqnarray}
&& x(t)=\frac{U^{\frac{1}{\alpha}}e^{-\frac{1}{\alpha}\int b_1dt}}
{\left[1+\frac{mq}{\beta}\displaystyle\int \left[a_1\left(Ue^{-\int b_1dt}\right)^{\frac{q}{\beta}}+a_2\left(Ve^{-\int b_2dt}\right)^{\frac{q}{\beta}}\right]^mdt\right]^{\frac{\beta}{\alpha qm}}},\\
&& y(t)=\frac{V^{\frac{1}{\beta}}e^{-\frac{1}{\beta}\int b_2dt}}
{\left[1+\frac{mq}{\beta}\displaystyle\int  \left[a_1\left(Ue^{-\int b_1dt}\right)^{\frac{q}{\beta}}+a_2\left(Ve^{-\int b_2dt}\right)^{\frac{q}{\beta}}\right]^mdt\right]^{\frac{1}{qm}}}.
\end{eqnarray}
\end{subequations}
Note that the general solution contains four arbitrary constants through $U$ and $V$, see Eq. (\ref{2linsol}).

As a specific example of (\ref{bernoulli-derived}), let us consider for the simple choice $\alpha=\beta=p=q=m=1$ and $a_1,\,a_2,\,b_1,\,b_2$ are constants.  For this
parametric choice Eq. (\ref{bernoulli-derived}) reduces to
\begin{eqnarray}
&& \ddot{x}+2(a_1x+a_2y+b_1)\dot{x}+(a_1\dot{x}+a_2\dot{y})x
+(a_1x+a_2y+b_1)^2x\nonumber\\
 &&\qquad\qquad\qquad\qquad\qquad\qquad\qquad
+c_{11}(\dot{x}+x(a_1x+a_2y+b_1))+c_{12}x=0,\nonumber\\
 &&\ddot{y}+2(a_1x+a_2y+b_2)\dot{y}+(a_1\dot{x}+a_2\dot{y})y
+(a_1x+a_2y+b_2)^2y\nonumber\\
 &&\qquad\qquad\qquad\qquad\qquad\qquad\qquad
+c_{21}(\dot{y}+y(a_1x+a_2y+b_2))+c_{22}y=0.
\label {case1}
\end{eqnarray}
 The general solution of Eq. (\ref{case1}) reads
\begin{eqnarray}
x(t)=\frac{Ue^{-b_1t}}
{1+\int (a_1Ue^{-b_1t}
+a_2Ve^{- b_2t})dt},\nonumber\\
y(t)=\frac{Ve^{- b_2t}}
{1+\int (a_1Ue^{- b_1t}
+a_2Ve^{-b_2t})dt}.
\label {lv071b}
\end{eqnarray}
 As one expects
the general solution of the nonlinear system depends upon the solution of the damped linear harmonic
oscillator equation too. As a consequence the solution of Eq. (\ref{case1}) may be periodic or decaying or growing type
depending upon the nature of the parameters, $c_{11},\,c_{12},\,c_{21}$, and $c_{22}$ which appear in (\ref{slv01}).  We note that for the parametric choice $b_1,\,b_2,\,c_{11},\,c_{21}=0$, Eq. (\ref{case1}) becomes Eq. (\ref{secondordercmee}) and admits isochronous oscillations where the frequency is independent of amplitude or initial conditions {\footnotesize$^{11,12}$}, see Eq. (\ref{cmee04b}).  Also note that for the choice $c_{11}=c_{12}=c_{21}=c_{22}=0$ and $b_1=b_2=0$, Eq. (\ref{case1}) reduces to the third member of the two coupled Ricatti chain (\ref{cmee}).

\noindent{\bf\underline{Case : 2}}

\noindent Apart from the above discussed form of $f$ and $g$ which reduces
Eq. (\ref{generalriccati}) to an integrable system of coupled Bernoulli type equations, we find the following form,
$\displaystyle f=\left(\sum_{i=1}^Na_i(t)x^{p_{i}}y^{q_{i}}\right)^m+b_1(t)$ and $\displaystyle g=\left(\sum_{i=1}^Na_i(t)x^{p_{i}}y^{q_{i}}\right)^m+b_2(t)$, where $p_i,\,q_i,\,i=1,2,\ldots,N$, and $m$ are arbitrary real constants, also reduces Eq. (\ref{generalriccati}) to the following coupled Bernoulli type equations
\begin{eqnarray}
\FL \dot{x}=\frac{x}{\alpha}
\left[\frac{\dot{U}}{U}-b_1(t)-\left(\sum_{i=1}^Na_i(t)x^{p_{i}}y^{q_{i}}\right)^m\right],\,
\dot{y}=\frac{y}{\beta}
\left[\frac{\dot{V}}{V}-b_2(t)-\left(\sum_{i=1}^Na_i(t)x^{p_{i}}y^{q_{i}}\right)^m\right].\label{coup-bern-eq1}
\end{eqnarray}
However we find that the general solution of Eq. (\ref{coup-bern-eq1}) can be deduced for the specific parametric restrictions $\frac{p_i\beta}{\alpha}+q_i$= constant, where $i=1,2,\ldots,N$.
Substituting this form of $f$ and $g$ in Eq. (\ref{gen-non-coup}) we get the associated integrable coupled
nonlinear second order ODEs,
\begin{subequations}
\begin{eqnarray}
\addtocounter{equation}{-1}
\label{second-coup-case2}
\addtocounter{equation}{1}
&& \ddot{x}+(\alpha-1)\frac{\dot{x}^2}{x}+2\dot{x}\left((\sum_{i=1}^Na_i x^{p_i}y^{q_i})^m+b_1+c_{11}\right)+\frac{x}{\alpha}\left[\bigg[\left(\sum_{i=1}^Na_ix^{p_i}y^{q_i}\right)^m+b_1\bigg]^2+c_{12}\right.\nonumber\\
&&\qquad +c_{11}\left((\sum_{i=1}^Na_ix^{p_i}y^{q_i})^m+b_1\right)
+m(\sum_{i=1}^Na_ix^{p_i}y^{q_i})^{m-1}\bigg[\sum_{i=1}^N(a_ip_ix^{p_i-1}y^{q_i}\dot{x}+a_iq_ix^{p_i}y^{q_i-1}\dot{y}\nonumber\\
&&\hspace{5cm}\left.+a_{it}x^{p_i}y^{q_i})\bigg]+b_{1t}\right]=0,\label{second-coup-case2a}
\end{eqnarray}
\begin{eqnarray}
&& \ddot{y}+(\beta-1)\frac{\dot{y}^2}{y}+2\dot{y}\left((\sum_{i=1}^Na_i x^{p_i}y^{q_i})^m+b_2\right)+\frac{y}{\beta}\left[\bigg[\left(\sum_{i=1}^Na_ix^{p_i}y^{q_i}\right)^m+b_2\bigg]^2+c_{22}\right.\nonumber\\
&&\qquad +c_{21}\left((\sum_{i=1}^Na_ix^{p_i}y^{q_i})^m+b_2\right)
+m(\sum_{i=1}^Na_ix^{p_i}y^{q_i})^{m-1}\bigg[\sum_{i=1}^N(a_ip_ix^{p_i-1}y^{q_i}\dot{x}+a_iq_ix^{p_i}y^{q_i-1}\dot{y}\nonumber\\
&&\hspace{5cm}\left.+a_{it}x^{p_i}y^{q_i})\bigg]+b_{2t}\right]=0.\label{second-coup-case2b}
\end{eqnarray}
\end{subequations}
 Again, multiplying the first equation of (\ref{coup-bern-eq1}) by $\alpha y$ and the second of (\ref{coup-bern-eq1}) by
 $\beta x$ and subtracting the latter from the former one obtains an equation for $(\alpha y\dot{x}-\beta x\dot{y})$ in exactly the same form as that of Eq. (\ref{coup-bern-1}).
 As before, dividing the resultant equation on both sides by $xy$ and integrating one gets
$x(t)=\left(\frac{U}{V}e^{\int(b_2-b_1)dt}y^{\beta}\right)^{\frac{1}{\alpha}}$.
Substituting this in the second equation of Eq. (\ref{coup-bern-eq1}) we get
\begin{eqnarray}
\dot{y}=\frac{1}{\beta}\left[\left(\frac{\dot{V}}{V}-b_2\right)y-\left(\sum_{i=1}^Na_i
\left(\frac{U}{V}e^{\int(b_2-b_1)dt}\right)^{\frac{p_i}{\alpha}}
y^{\frac{p_i\beta}{\alpha}+q_i+1}\right)^m\right].\label{bern1}
\end{eqnarray}
The above equation reduces to the Bernoulli type equation (\ref{bernoulli-type}) for the following parametric restrictions
\begin{eqnarray}
\frac{p_i\beta}{\alpha}+q_i+1=\frac{k}{m},\,\,i=1,2,\ldots,N,\label{cond1}
\end{eqnarray}
where $k$ is a constant.  The parametric restrictions (\ref{cond1}) contains $N$ equations as $i$ runs from $1$ to $N$ and this set of conditions implies that all the terms in the expression $\displaystyle\sum_{i=i}^Na_ix^{p_i}y^{q_i}$ are of the same degree.
The parametric restrictions (\ref{cond1}) reduce equation (\ref{bern1}) to
\begin{eqnarray}
\dot{y}=\frac{1}{\beta}\left[\left(\frac{\dot{V}}{V}-b_2\right)y-\left(\sum_{i=1}^Na_i
\left(\frac{U}{V}e^{\int(b_2-b_1)dt}\right)^{\frac{p_i}{\alpha}}\right)^my^k\right].
\end{eqnarray}
 Integrating equation (\ref{bern1}) we obtain $y(t)$ and once $y$ is known, $x$ can be
deduced straightforwardly through (\ref{xyrelation}).  Doing so one obtains
\begin{subequations}
\addtocounter{equation}{-1}
\label{gen-sol3-gen}
\addtocounter{equation}{1}
\begin{eqnarray}
&&\qquad x(t)=\frac{U^{\frac{1}{\alpha}}e^{-\frac{1}{\alpha}\int b_1 dt}}
{\left[1+\frac{k-1}{\beta}\displaystyle\int V^{\frac{(k-1)}{\beta}}e^{\frac{(1-k)}{\beta}\int b_2dt}\left(
\sum_{i=1}^N a_i\left(\frac{U}{V}e^{\int(b_2-b_1)dt}\right)^{\frac{p_i}{\alpha}}\right)^mdt\right]^{\frac{\beta}{\alpha(k-1)}}},
\\
&&\qquad y(t)=\frac{V^{\frac{1}{\beta}}e^{-\frac{1}{\beta}\int b_2dt}}
{\left[1+\frac{k-1}{\beta}\displaystyle\int V^{\frac{(k-1)}{\beta}}e^{\frac{(1-k)}{\beta}\int b_2dt}\left(
\sum_{i=1}^N  a_i\left(\frac{U}{V}e^{\int(b_2-b_1)dt}\right)^{\frac{p_i}{\alpha}}\right)^mdt\right]^{\frac{1}{k-1}}}.
\end{eqnarray}
\end{subequations}
As an interesting example, for the choice $\alpha=\beta=m=p_i=q_i=N=1$, Eq. (\ref{second-coup-case2}) reduces to
\begin{subequations}
\begin{eqnarray}
\addtocounter{equation}{-1}
\label{eq34}
\addtocounter{equation}{1}
&& \ddot{x}+2\dot{x}(a_1xy+b_1+c_{11})+a_1x^2\dot{y}+a_1^2x^3y^2+x^2y(a_{1t}+2a_1b_1)\nonumber\\
&&\hspace{5cm}+x(b_1c_{11}+b_{1t}+c_{12}+b_1^2)=0,\\
&& \ddot{y}+2\dot{y}(a_1xy+b_2+c_{21})+a_1x^2\dot{y}+a_1^2x^3y^2+x^2y(a_{1t}+2a_1b_2)\nonumber\\
&&\hspace{5cm}+x(b_2c_{21}+b_{2t}+c_{22}+b_2^2)=0.
\end{eqnarray}
\end{subequations}
The general solution of the above equation is
\begin{subequations}
\begin{eqnarray}
\addtocounter{equation}{-1}
\label{sol34}
\addtocounter{equation}{1}
 x(t)=\frac{Ue^{-\int b_1 dt}}
{\left[1+2\int a_1 UV
e^{-\int(b_2+b_1)dt}dt\right]^{\frac{1}{2}}},\label{sol33}\\
 y(t)=\frac{Ve^{-\int b_2 dt}}
{\left[1+2\int a_1 UV
e^{-\int(b_2+b_1)dt}dt\right]^{\frac{1}{2}}}.
\end{eqnarray}
\end{subequations}
Note that solution (\ref{sol34}) can become isochronous for the choice $b_1=b_2=0$ and suitable choice of other
parameters in (\ref{eq34}).  Also note that for the choice $c_{11}=c_{12}=c_{21}=c_{22}=0$, $b_1=b_2=0$ and $a_1,\,a_2$ are constants, Eq. (\ref{eq34}) reduces to the third member of the new integrable chain (\ref{newchain3}).

\section{Integrable Arbitrary $n$-coupled second order nonlinear ODEs}
\label{section5}
The above analysis can be generalized in principle without much difficulty to a system of arbitrary $n$-coupled second order ODEs and classes of solvable ones from the linear ODEs can be identified.  In order to do so we first investigated the case of three-coupled second order ODEs.  Then from the results of two- and three-coupled second order ODEs, we have generalized the results to $n$-coupled second order ODEs.   For brevity, we are not presenting here the results of the three-coupled second order ODEs (one can refer to Ref.[17] for details), but straightaway present the results for a system of $n$-coupled ODEs.

Let us consider a system of $n$ uncoupled linear second order ODEs of the form
\begin{eqnarray}
\ddot{U}_i+c_{i1}\dot{U}_i+c_{i2} U_i=0,\,\,\,i=1,...,n,
\label {nv04}
\end{eqnarray}
whose general solution can be expressed as $U_i(t)=I_{1i}e^{m_{1i}t}+I_{2i}e^{m_{2i}t}$, where $I_{1i}$
and $I_{2i}$, $i=1,2,\ldots,n$, are arbitrary constants and $m_{1i}$'s and $m_{2i}$'s are the solutions of the
auxiliary equations.

\noindent{\underline{\bf Case : 1}}

\noindent Now introducing the nonlocal transformations analogous to (\ref{form-f1})
\begin{eqnarray}
U_i=x_i^{\alpha_i}e^{\int((\sum_j^na_{j}(t)x_j^{p_j})^m+b_i(t))dt'},\,\,i=1,...,n,\label{nv01}
\end{eqnarray}
in (\ref{nv04}), one can transform the uncoupled damped linear harmonic oscillators
into the sytem of
 $n$-coupled nonlinear second order ODEs of the form
\begin{eqnarray}
&&\ddot{x}_i+(\alpha_i-1)\frac{\dot{x}_i^2}{x_i}
+2((\sum_j^n a_{j}x_j^{p_j})^m+b_i)\dot{x}_i
+\frac{mx_i}{\alpha_i}\bigg[(\sum_{i=1}^na_jx^{p_j})^{m-1}\bigg(\sum_{i=1}^n
a_jp_jx^{p_j-1}\dot{x}_j+a_{jt}x^{p_j}\bigg)\bigg]\nonumber\\
&& \qquad+\frac{x_i}{\alpha_i}\bigg[\bigg((\sum_{i=1}^na_jx_j^{p_j})^m+b_i\bigg)^2
+c_{i1}((\sum_{i=1}^na_jx_j^{p_j})^m+b_i)+c_{i2}\bigg]+c_{i1}\dot{x}
=0,\,i=1,2,...n.
\label {nnv02}
\end{eqnarray}

Now we confine our attention on obtaining the general solution for Eq. (\ref{nnv02}) which can be done
by following the same methodology described earlier for the two coupled case.  Then
 Eqs. (\ref{nv01})
can be brought to the form of a set of coupled Bernoulli type first order ODEs,
\begin{eqnarray}
\dot{x}_i=\frac{\dot{U}_i}{U_i}x_i-((\sum_j^n a_{j}x_j^{p_j})^m+b_i)x_i,\,\,i=1,...,n.
\label {nv07}
\end{eqnarray}
One can reduce the above system of $n$-coupled first order ODEs to a single first order equation as was done in the case of $n=2$ (vide Eq. \ref{bern2}).  For a better understanding we demonstrate the procedure for $n=3$.
For this case the above system of equations (\ref{nv07}) reduces to
\begin{subequations}
\begin{eqnarray}
\addtocounter{equation}{-1}
\label{coup-bernoulli-3}
\addtocounter{equation}{1}
\FL\qquad\qquad\dot{x}_1
=\frac{x_1}{\alpha_1}\left[\frac{\dot{U}_1}{U_1}-b_1-(a_{1}x^{p_1}+a_{2}x_2^{p_2}+a_3x_3^{p_3})^m\right],\label{coup-bernoulli-3a}\\
\FL\qquad\qquad \dot{x}_2=\frac{x_2}{\alpha_2}\left[\frac{\dot{U}_2}{U_2}-b_2-(a_{1}x_1^{p_1}+a_{2}x_2^{p_2}+a_3x_3^{p_3})^m\right],\label{coup-bernoulli-3b}
\\
\FL \qquad\qquad \dot{x}_3=\frac{x_3}{\alpha_3}
\left[\frac{\dot{U}_3}{U_3}-b_3
-(a_{1}x_1^{p_3}+a_{2}x_2^{p_2}+a_3x_3^{p_3})^m\right].\label{coup-bernoulli-3c}
\end{eqnarray}
\end{subequations}
We solve the above system of equations using the same method used to obtain (\ref{xyrelation}).
 Multiplying (\ref{coup-bernoulli-3a}) by $\alpha_1 x_3$ and
(\ref{coup-bernoulli-3c}) by $\alpha_3 x_1$ and subtracting the latter from the former one obtains
\begin{eqnarray}
\alpha_1 x_3\dot{x}_1-\alpha_3 x_1\dot{x}_3=\left(\frac{\dot{U}_1}{U_1}-\frac{\dot{U}_3}{U_3}+b_3(t)-b_1(t)\right)x_1x_3.
\label{third-coup-bern1}
\end{eqnarray}
 Dividing both sides by $x_1x_3$ and integrating the resultant equation (\ref{third-coup-bern1}), after absorbing the constant of integration, one gets
\begin{eqnarray}
x_1=\left(\frac{U_1}{U_3}e^{\int(b_3-b_1)dt}x_3^{\alpha_3}\right)^{\frac{1}{\alpha_1}}.\label{xzrelation}
\end{eqnarray}
Multiplying (\ref{coup-bernoulli-3b}) by $\alpha_2 x_3$ and
(\ref{coup-bernoulli-3c}) by $\alpha_3 x_2$ and subtracting the latter from the former one obtains
\begin{eqnarray}
\alpha_2 x_3\dot{x}_2-\alpha x_2\dot{x}_3=\left(\frac{\dot{U}_2}{U_2}-\frac{\dot{U}_3}{U_3}+b_3(t)-b_2(t)\right)x_2x_3.
\label{third-coup-bern2}
\end{eqnarray}
Dividing both sides by $x_2x_3$ and integrating the resultant equation (\ref{third-coup-bern2}) one gets
\begin{eqnarray}
x_2=\left(\frac{U_2}{U_3}e^{\int(b_3-b_2)dt}x_3^{\alpha_3}\right)^{\frac{1}{\alpha_2}}.\label{yzrelation}
\end{eqnarray}
Substituting (\ref{xzrelation}) and (\ref{yzrelation}) in
 Eq. (\ref{coup-bernoulli-3c}) we get
\begin{eqnarray}
\FL \dot{x}_3=\frac{x_3}{\alpha_3}\left[\frac{\dot{U}_3}{U_3}-b_3-
\left\{a_1\left(\frac{U_1}{U_3}e^{\int(b_3-b_1)dt}\right)^{\frac{p_1}{\alpha_1}}x_3^{\frac{p_1\alpha_3}{\alpha_1}}
+a_2\left(\frac{U_2}{U_3}e^{\int(b_3-b_2)dt}\right)^{\frac{p_2}{\alpha_2}}
x_3^{\frac{p_2\alpha_3}{\alpha_2}}+a_3x_3^{p_3}\right\}^m\right].\label{third-bern}
\end{eqnarray}
In a similar way the system of $n$ coupled Bernoulli type equations (\ref{nv07}) can be brought to the single first order equation for the $n$th variable $x_n(t)$,
\begin{eqnarray}
\FL\qquad\quad \dot{x}_n=\frac{x_n}{\alpha_n}\left[\frac{\dot{U}_n}{U_n}-b_n-\left[\sum_{i=1}^{n}a_i\left(\frac{U_i}{U_n}e^{\int(b_n-b_i)dt}\right)^{\frac{p_i}{\alpha_i}}x_n^{\frac{p_i\alpha_n}{\alpha_i}}\right]^m\right].
\label{ncoupled-bern}
\end{eqnarray}

Utilizing the same idea given in the previous case and using the parametric restriction $\frac{p_1\alpha_n}{\alpha_1}=\frac{p_2\alpha_n}{\alpha_2}=...=p_n$, one can inductively deduce the general solution of Eq. (\ref{nnv02}) as
\begin{eqnarray}
\FL\qquad\quad x_i(t)=\frac{U_i^{\frac{1}{\alpha_i}}e^{-\frac{1}{\alpha_i}\int b_idt}}{\left[1+\frac{mp_n}{\alpha_n}\displaystyle\int \left[\sum_{j=1}^{n}a_j\left(U_je^{-\int b_jdt}\right)^{\frac{p_n}{\alpha_n}}\right]^mdt \right]^{\frac{\alpha_n}{mp_n\alpha_i}}},\,\,i=1,2,\ldots,n.
\label {nv08}
\end{eqnarray}
The general solution contains $2n$ arbitrary constants, coming from that of solutions of Eq. (\ref{nv04}).

\noindent{\underline{\bf Case : 2}} 

\noindent Substituting the nonlocal transformation
$U_i=x_i^{\alpha_i}e^{\int (\sum_{j=1}^Na_j(t)\prod_{k=1}^nx_{k}^{p_{jk}})^m+b_i(t)}$, in Eq. (\ref{nv04}) we get the set of $n$ coupled second order nonlinear ODEs,
\begin{eqnarray}
&& \ddot{x}_i+(\alpha_i-1)\frac{\dot{x}_i^2}{x_i}+2\dot{x}_i\left((\sum_{j=1}^Na_j\prod_{k=1}^nx_{k}^{p_{jk}})^m+b_i\right)+\frac{m}{\alpha}\left[\sum_{j=1}^N\left( a_j\sum_{k=1}^n p_k x_k^{p_k-1}\dot{x}_k
\prod_{l=1\atop l\ne k}^n x_l^{p_l}\right)\right.\nonumber\\
&&\quad\left.+\sum_{j=1}^Na_{jt}\prod_{k=1}^nx_{k}^{p_{jk}}\right](\sum_{j=1}^Na_j\prod_{k=1}^n x_k^{p_{jk}})^{m-1}
+\frac{c_{i1}x_i}{\alpha_i}\left[(\sum_{j=1}^Na_j\prod_{k=1}^nx_{k}^{p_{jk}})^m+b_i\right]+\frac{x_i}{\alpha_i}(c_{i2}+b_{it})\nonumber\\
&&\qquad\qquad+\frac{x_i}{\alpha_i}\left[ (\sum_{j=1}^Na_j\prod_{k=1}^nx_{k}^{p_{jk}})^m+b_i\right]^2+c_{i1}\dot{x}=0.
\end{eqnarray}
Proceeding as before, for the parametric choice $\displaystyle\sum_{i=1}^n \frac{p_{ij}\alpha_n}{\alpha_i}+1=\frac{\kappa}{m},\,\,\,j=1,2,\ldots,N$, the solution of the above equation is given by
\begin{eqnarray}
x_i(t)=\frac{U_i^{\frac{1}{\alpha_i}}e^{-\frac{1}{\alpha_i}\int b_i dt}}
{\left[1+\frac{\kappa-1}{\alpha_n}\displaystyle\int s(t) (U_ne^{-\int b_n dt})^{\frac{(\kappa-1)}{\alpha_n}}dt\right]^{\frac{\alpha_n}{\alpha_i(\kappa-1)}}},\label{nsecondsol}
\end{eqnarray}
where
$s(t)=\left[\sum_{i=1}^N a_i \prod_{j=1}^{n-1} \left[\frac{U_j}{U_n}\right]^{\frac{p_j}{\alpha_j}}e^{\frac{p_j}{\alpha_j}\int(b_n-b_j)dt}\right]^m$. Here the set $U_i$, $i=1,2,3,\ldots,n,$ is the solution of the linear system of ODEs (\ref{nv04}).

\section{Integrable Coupled third order nonlinear ODEs}
\label{section6}
\subsection{Two coupled third order ODEs}
Having studied the nonlocal connection between second order linear and nonlinear ODEs now we generalize the results to a system of third order ODEs.  For this purpose let us consider the
following system of uncoupled third order linear ODEs,
\begin{eqnarray}
\dddot{U}+c_{11}\ddot{U}+c_{12} \dot{U}+c_{13} U=0,\quad
\dddot{V}+c_{21}\ddot{V}+c_{22} \dot{V}+c_{23} V=0,
\label {tlv01}
\end{eqnarray}
where $c_{i,j}$ $i=1,2$, $j=1,2,3,$ are arbitrary parameters, and the same nonlocal
transformation (\ref{nonlocal-trans}),
\begin{eqnarray}
U=x^\alpha e^{\int f(x,y,t) dt},\,\,V=y^\beta e^{\int g(x,y,t) dt}.\nonumber
\end{eqnarray}
Substituting the above nonlocal transformation
 in
(\ref{tlv01}) we obtain the following system of two coupled third order nonlinear ODEs,
\begin{subequations}
\begin{eqnarray}
\addtocounter{equation}{-1}
\label{gen-third-2coup}
\addtocounter{equation}{1}
&&\dddot{x}+(3(\alpha-1)\frac{\dot{x}}{x}+3f+c_{11}+\frac{x}{\alpha}f_x)\ddot{x}+\frac{x}{\alpha}f_y\ddot{y}+(\alpha-1)(\alpha-2)\frac{\dot{x}^3}{x^2}+(\frac{(\alpha-1)}{x}(3f+c_{11})+3f_x\nonumber\\
&&\quad+\frac{x}{\alpha}f_{xx})\dot{x}^2+(3f_y+\frac{2x}{\alpha}f_{xy})\dot{x}\dot{y}+(3f^2+2c_{11}f+c_{12}+\frac{2x}{\alpha}f_{xt}+\frac{3x}{\alpha}ff_x+\frac{c_{11}x}{\alpha}f_x+3f_t)\dot{x}\nonumber\\
&&\quad+(\frac{2x}{\alpha}f_{yt}+\frac{3x}{\alpha}ff_y+\frac{c_{11}x}{\alpha}f_y)\dot{y}+\frac{x}{\alpha}(3ff_t+f^3+c_{11}f^2+c_{11}f_t+f_{tt}+c_{13}+c_{12}f)=0,
\end{eqnarray}
\begin{eqnarray}
&&\dddot{y}+(3(\beta-1)\frac{\dot{y}}{y}+3g+c_{21}+\frac{y}{\beta}g_y)\ddot{y}+\frac{y}{\alpha}g_x\ddot{x}+(\beta-1)(\beta-2)\frac{\dot{y}^3}{y^2}+(\frac{(\beta-1)}{y}(3g+c_{21})+3g_y\nonumber\\
&&\quad+\frac{y}{\beta}g_{yy})\dot{y}^2+(3g_x+\frac{2y}{\beta}g_{xy})\dot{x}\dot{y}+(3g^2+2c_{21}g+c_{22}+\frac{2y}{\beta}g_{yt}+\frac{3y}{\beta}gg_y+\frac{c_{21}y}{\beta}g_y+3g_t)\dot{y}\nonumber\\
&&\quad+(\frac{2y}{\beta}f_{xt}+\frac{3y}{\beta}gg_x+\frac{c_{21}y}{\beta}g_x)\dot{x}+\frac{y}{\beta}(3gg_t+g^3+c_{11}g^2+c_{21}g_t+g_{tt}+c_{23}+c_{22}g)=0.
\end{eqnarray}
\end{subequations}
As in the case of the system of second order ODEs, we find that there are two sets of $f$ and $g$ (vide Eqs. (\ref{form-f1}) and (\ref{form-g1})),
for which the above equation becomes integrable.  We briefly discuss below the general solution
of Eqs. (\ref{gen-third-2coup}) for both the choices.

\noindent{\bf\underline{Case : 1}}

\noindent Substituting $f=(a_1(t)x^p+a_2(t)y^q)^m+b_1(t)$ and
$g=(a_1(t)x^p+a_2(t)y^q)^m+b_2(t)$
in (\ref{gen-third-2coup}) we get a system of two coupled third order
nonlinear ODEs. The solution of this set of nonlinear ODEs can again be obtained by solving the following
coupled Bernoulli type equations,
\begin{subequations}
\begin{eqnarray}
\addtocounter{equation}{-1}
\label{third-coup-bernoulli-1}
\addtocounter{equation}{1}
&&\dot{x}=\frac{x}{\alpha}\left[\frac{\dot{U}}{U}-b_1(t)-(a_{1}(t)x^{p}+a_{2}(t)y^{q})^m\right],\\
&&\dot{y}=\frac{y}{\beta}\left[\frac{\dot{V}}{V}-b_2(t)-(a_{1}(t)x^{p}+a_{2}(t)y^{q})^m\right].
\end{eqnarray}
\end{subequations}
which is obtained through the nonlocal transformation $U=x^{\alpha}e^{\int((a_1(t)x^p+a_2(t)y^q)^m+b_1(t))dt}$, $V=y^{\alpha}e^{\int ((a_1(t)x^p+a_2(t)y^q)^m+b_2(t))dt}$ in (\ref{gen-third-2coup}) with the above forms of $f$ and $g$.

We find that the above equation (\ref{third-coup-bernoulli-1}) is exactly the same as that of equation (\ref{coup-bernoulli-1}) whose solution has been discussed in Sec \ref{section3}, except for the difference that $U$ and $V$ are now solutions of the third order linear ODEs (\ref{tlv01}).  Therefore we directly write down the solution of equation (\ref{third-coup-bernoulli-1}) (vide Eq. (\ref{gen-sol1-gen}))  as
\begin{subequations}
\addtocounter{equation}{-1}
\label{third-gensol}
\addtocounter{equation}{1}
\begin{eqnarray}
\FL\qquad\quad x(t)=\frac{U^{\frac{1}{\alpha}}e^{-\frac{1}{\alpha}\int b_1dt}}{\left[1+\frac{mq}{\beta}\displaystyle\int \left[a_1\left(Ue^{-\int b_1dt}\right)^{\frac{q}{\beta}}+a_2\left(Ve^{-\int b_2dt}\right)^{\frac{q}{\beta}}\right]^m dt\right]^{\frac{\beta}{\alpha mq}}},\\
\FL\qquad\quad y(t)=\frac{V^{\frac{1}{\beta}}e^{-\frac{1}{\beta}\int b_2dt}}{\left[1+\frac{mq}{\beta}\displaystyle\int \left[a_1\left(Ue^{-\int b_1)dt}\right)^{\frac{q}{\beta}}+a_2\left(Ve^{-\int b_2)dt}\right)^{\frac{q}{\beta}}\right]^m dt \right]^{\frac{1}{ mq}}},
\end{eqnarray}
\end{subequations}
where $U$ and $V$ are now the general solutions of the system of third order linear ODEs (\ref{tlv01}), which can be given in the form 
\begin{eqnarray}
U(t)=I_1e^{m_1t}+I_2e^{m_2t}+I_3e^{m_3t},\quad V(t)=I_4e^{m_4t}+I_5e^{m_5t}+I_6e^{m_6t},
\end{eqnarray}
where $I_1,I_2,I_3,\ldots,I_6$, are six arbitrary constants and $m_1,m_2,m_3$ are the solutions of the auxiliary equation 
\begin{subequations}
\addtocounter{equation}{-1}
\label{eigenequations}
\addtocounter{equation}{1}
\begin{eqnarray}
m^3+c_{11}m^2+c_{12}m+c_{13}=0, 
\end{eqnarray}
while $m_4,m_5,m_6$ are the solutions of the auxiliary equation 
\begin{eqnarray}
m^3+c_{21}m^2+c_{22}m+c_{23}=0.
\end{eqnarray}
\end{subequations}
We note that for the parametric choice $c_{11}=c_{13}=c_{21}=c_{23}=0$ the general solution of Eq. (\ref{tlv01}), $U$ and $V$, become periodic and for this parametric choice with $b_1=b_2=0$, we find that the solution (\ref{third-gensol}) exhibits isochronous behaviour.
As specific example let us consider the choice $\alpha=\beta=p=q=1$, $b_1=b_2=0$, $a_1,\,a_2$ constants. Then the nonlocal transformation $U=x^{\alpha}e^{\int (a_1(t)x^p+a_2(t)y^q)^m+b_1(t)dt}$, $V=y^{\alpha}e^{\int (a_1(t)x^p+a_2(t)y^q)^m+b_2(t)dt}$ reduces to the form $U=xe^{\int (a_1x+a_2y)dt}$ and $V=ye^{\int (a_1x+a_2y)dt}$.  Substituting this form of nonlocal transformation in
\begin{eqnarray}
\dddot{U}=0,\qquad\dddot{V}=0
\end{eqnarray}
we get the following set of third order nonlinear ODEs,
\begin{eqnarray}
\FL &&\quad \dddot{x}+3(a_1\dot{x}+a_2\dot{y})\dot{x}
+2(a_1x+a_2y)\ddot{x}+(a_1\ddot{x}+a_2\ddot{y})x+2(a_1\dot{x}+a_2\dot{y})x
+(a_1x+a_2y)^2\dot{x}\nonumber\\
\FL &&\qquad+(a_1x+a_2y)(\ddot{x}+2(a_1x+a_2y)\dot{x}
+(a_1\dot{x}+a_2\dot{y})x+(a_1x+a_2y)^2x)=0,\nonumber\\
\FL &&\quad \dddot{y}+3(a_1\dot{x}+a_2\dot{y})\dot{y}
+2(a_1x+a_2y)\ddot{y}+(a_1\ddot{x}+a_2\ddot{y})y+2(a_1\dot{x}+a_2\dot{y})y
+(a_1x+a_2y)^2\dot{y}\nonumber\\
\FL &&\qquad+(a_1x+a_2y)
(\ddot{y}+2(a_1x+a_2y)\dot{y}+(a_1\dot{x}+a_2\dot{y})y
+(a_1x+a_2y)^2y)=0.
\label {tfeq01}
\end{eqnarray}
Equation (\ref{tfeq01}) is nothing but a system of two coupled Chazy equation XII
(with $N=2$ and parametric restrictions $A = 0, B = 0$ in
Ref. [18]), which has been studied in detail in
Refs. [18-23]. 

\noindent{\bf\underline{Case : 2}} We substitute $\displaystyle f=\left(\sum_{i=1}^Na_i(t)x^{p_{i}}y^{q_{i}}\right)^m+b_1(t)$ and $\displaystyle g=\left(\sum_{i=1}^Na_i(t)x^{p_{i}}y^{q_{i}}\right)^m+b_2(t)$
(which is the same as used earlier in Sec. \ref{section3} for second order ODEs) to obtain
\begin{subequations}
\begin{eqnarray}
\addtocounter{equation}{-1}
\label{third-coup-case2}
\addtocounter{equation}{1}
\FL\dddot{x}+\delta_1(x,y,t)\ddot{x}+\delta_2(x,y,t)\ddot{y}+(\alpha-1)(\alpha-2)\frac{\dot{x}^3}{x^2}+\delta_3(x,y,t)\dot{x}^2+\delta_4(x,y,t)\dot{y}^2\nonumber\\
+\delta_5(x,y,t)\dot{x}\dot{y}+\delta_6(x,y,t)\dot{x}+\delta_7(x,y,t)\dot{y}+\delta_8(x,y,t)=0,\\
\FL\dddot{y}+\epsilon_1(x,y,t)\ddot{x}+\epsilon_2(x,y,t)\ddot{y}+(\beta-1)(\beta-2)\frac{\dot{y}^3}{y^2}+\epsilon_3(x,y,t)\dot{x}^2+\epsilon_4(x,y,t)\dot{y}^2\nonumber\\
+\epsilon_5(x,y,t)\dot{x}\dot{y}+\epsilon_6(x,y,t)\dot{x}+\epsilon_7(x,y,t)\dot{y}+\epsilon_8(x,y,t)=0,
\end{eqnarray}
\end{subequations}
where $\delta_i(x,y,t)$ and $\epsilon_i(x,y,t)$, $i=1,2,...,8$, are  functions of the indicated variables and we do not explicitly present them here as they are complicated expressions and one may refer to Ref. [17] for details.

As we have pointed out in the previous case, the solution of the two-coupled second order nonlinear equation (\ref{second-coup-case2}) and the solution of the two-coupled third order nonlinear equation (\ref{third-coup-case2}) are the same except for the form of the solutions of the linear ODEs.  Therefore one can write down the solution of Eq. (\ref{third-coup-case2}) also directly using the solution of the two-coupled second order nonlinear equation (\ref{second-coup-case2}) with the parametric restriction $\frac{p_i\beta}{\alpha}+q_i+1=\frac{k}{m},\,\,i=1,2,\ldots,N$, as
\begin{subequations}
\begin{eqnarray}
&&\qquad x(t)=\frac{U^{\frac{1}{\alpha}}e^{-\frac{1}{\alpha}\int b_1 dt}}
{\left[1+\frac{k-1}{\beta}\displaystyle\int V^{\frac{(k-1)}{\beta}}e^{\frac{(1-k)}{\beta}\int b_2dt}\left(
\sum_{i=1}^N  a_i\left(\frac{U}{V}e^{\int(b_2-b_1)dt}\right)^{\frac{p_i}{\alpha}}\right)^mdt\right]^{\frac{\beta}{\alpha(k-1)}}},
\\
&&\qquad y(t)=\frac{V^{\frac{1}{\beta}}e^{-\frac{1}{\beta}\int b_2dt}}
{\left[1+\frac{k-1}{\beta}\displaystyle\int V^{\frac{(k-1)}{\beta}}e^{\frac{(1-k)}{\beta}\int b_2dt}\left(
\sum_{i=1}^N  a_i\left(\frac{U}{V}e^{\int(b_2-b_1)dt}\right)^{\frac{p_i}{\alpha}}\right)^mdt\right]^{\frac{1}{k-1}}},
\end{eqnarray}
\end{subequations}
where $U$ and $V$ are now the general solutions of the system of third order linear ODEs (\ref{tlv01}) given as in Eq. (\ref{eigenequations}).

\subsection{Arbitrary $n$-coupled third order ODEs}
We conclude this section by extending the ideas to a sytem $n$-coupled third order nonlinear ODEs
by extending the ideas of the previous discussion. Let us consider a system of $n$ uncoupled linear third order ODEs of the
form
\begin{eqnarray}
\dddot{U}_i+c_{i1}\ddot{U}_i+c_{i2}\dot{U}_i+c_{i3}U_i=0,\,\,i=1,2,...,n.\label{linearnthird}
\end{eqnarray}
Now introducing the nonlocal transformation
\begin{eqnarray}
U_i=x_i^{\alpha_i}e^{\int f_i(x_1,\,x_2,\,...,\,x_n,\,t)dt},\quad i=1,2,\ldots,n,
\end{eqnarray}
one can derive the following set of coupled nonlinear ODEs,
\begin{eqnarray}
&& \dddot{x}_i+\left(3(\alpha_i-1)\frac{\dot{x}_i}{x_i}+3f+c_{i1}\right)\ddot{x}_i+\sum_{j=1}^n\frac{x_j\ddot{x}_j}{\alpha_i}+\left(\frac{(\alpha_i-1)}{x_i}(3f_i+c_{i1})+3f_{ix_i}\right)\dot{x}_i^2\nonumber\\
&&\quad+\sum_{j=1}^n\frac{x_jf_{ix_jx_j}}{\alpha_i}\dot{x}_j^2+(\alpha_i-1)(\alpha_i-2)\frac{\dot{x}^3_i}{x_i^2}+\sum_{l=1\atop l\ne i}^n3f_{ix_l}\dot{x}_i\dot{x}_l+\frac{2x_i}{\alpha_i}\sum_{j,l=1\atop j\ne l}^nf_{ix_jx_l}\dot{x}_j\dot{x}_l\nonumber\\
&&\quad+(3f_i^2+2c_{i1}f_i+c_{i2}+3f_{it})\dot{x}_i+\sum_{j=1}^n(\frac{2x_j}{\alpha_j}f_{x_jt}+\frac{3x_j}{\alpha_i}f_if_{ix_j}+\frac{c_{i1}x_j}{\alpha_i})\dot{x}_j+\frac{x_i}{\alpha_i}(3f_if_{it}+f_i^3\nonumber\\
&&\quad+c_{i1}f_i^2+c_{i1}f_{it}+f_{itt}+c_{i2}f_{i}+c_{i3})=0,\quad i=1,2,\ldots,n. \label{nthird}
\end{eqnarray}
We again consider two different forms of $f_i$'s for which one is able to deduce the solution of Eq. (\ref{nthird})
 separately.

\noindent\underline{\bf Case : 1}

Considering the form of $f_i$ as $f_i=(\sum_j^n a_{j}x_j^{p_j})^m+b_i)$
and imposing the parametric restrictions $\frac{p_1\alpha_n}{\alpha_1}=\frac{p_2\alpha_n}{\alpha_2}=...=p_n$, we obtain the following solution, as in the case of the coupled second order ODEs given by Eq. (\ref{nv08}),
\begin{eqnarray}
&&\qquad x_i(t)=\frac{U_i^{\frac{1}{\alpha_i}}e^{-\frac{1}{\alpha_i}\int b_idt}}
{\left[1+\frac{mp_n}{\alpha_n}\displaystyle\int
 \left[\sum_{j=1}^{n}a_j\left(U_je^{-\int b_jdt}\right)^{\frac{p_n}{\alpha_n}}\right]^mdt 
 \right]^{\frac{\alpha_n}{mp_n\alpha_i}}},\quad i=1,2,\ldots,n,
\end{eqnarray}
where $U_i$'s are the solutions of the system of third order linear ODEs (\ref{linearnthird}) containing $3N$ arbitrary constants.

\noindent\underline{\bf Case : 2}

Considering the form of $f_i$ as $f_i=(\sum_{j=1}^Na_j(t)\prod_{\kappa=1}^nx_{\kappa}^{p_{j\kappa}})^m+b_i(t)$ and imposing the parametric restrictions  $\displaystyle\sum_{i=1}^n \frac{p_{ij}\alpha_n}{\alpha_i}+1=\frac{k}{m}$, $j=1,2,\ldots,N$, we obtain the following solution
\begin{eqnarray}
&&\qquad x_i(t)=\frac{U_i^{\frac{1}{\alpha_i}}e^{-\frac{1}{\alpha_i}\int b_i dt}}
{\left[1+\frac{k-1}{\alpha_n}\displaystyle\int s(t) (U_Ne^{-\int b_n dt})^{\frac{(k-1)}{\alpha_n}}dt\right]^{\frac{\alpha_n}{\alpha_i(k-1)}}},\quad i=1,2,\ldots,n,\label{ngensol}
\end{eqnarray}
where
$s(t)=\left[\sum_{i=1}^N a_i \prod_{j=1}^{n-1}\left[\frac{U_j}{U_n}\right]^{\frac{p_j}{\alpha_j}}e^{\frac{p_j}{\alpha_j}\int(b_n-b_j)dt}\right]^m$, and $U_i$'s are the solutions of the system of third order linear ODEs (\ref{linearnthird}).  Again (\ref{ngensol}) is of the same form as (\ref{nsecondsol}) obtained for the case of coupled second order ODEs

\section{Integrable Coupled $l^{\mbox{th}}$ order nonlinear ODEs}
\label{section7}
Having studied the nonlocal connection that exists between linear and nonlinear
ODEs for the second and third orders we finally build a theory which is applicable for arbitrary $l^{\mbox{th}}$ order equations.

Consider a system of two uncoupled linear ODEs
\begin{eqnarray}
&&\bigg(\frac{d^l}{dt^l}+c_{11}\frac{d^{(l-1)}}{dt^{(l-1)}}
+\ldots+c_{1l-1}\frac{d}{dt}+c_{1l}\bigg)U(t)=0,\nonumber\\
&&\bigg(\frac{d^l}{dt^l}+c_{21}\frac{d^{(l-1)}}{dt^{(l-1)}}
+\ldots+c_{2l-1}\frac{d}{dt}+c_{2l}\bigg)V(t)=0, \label {nlv01}
\end{eqnarray}
where $c_{ij}$'s, $i=1,2,\;j=1,2,\ldots l$, are arbitrary constants.
The nonlocal
transformation (\ref{nonlocal-trans}) connects (\ref{nlv01}) to the set of coupled nonlinear
ODEs of the form
\begin{eqnarray}
&&\bigg(D_{1}^{(l)}+c_{11}D_{1}^{(l-1)}+\ldots+c_{1l-1}D_{1}^{(1)}
+c_{1l}\bigg)x=0,\nonumber\\
&&\bigg(D_{2}^{(l)}+c_{21}D_{2}^{(l-1)}+\ldots+c_{2l-1}D_{2}^{(1)}
+c_{2l}\bigg)y=0,
\label {nlv02}
\end{eqnarray}
where $D_{1}^{(l)} =\left(\alpha x^{\alpha-1}(\frac{d}{d t}+xf(x,y,t))\right)^l$ and
$D_{2}^{(l)} =\left(\beta y^{\beta-1}(\frac{d}{d t}+y g(x,y,t))\right)^l$.
As we have seen, the solution of Eq. (\ref{nlv02}) can be deduced from the nonlocal transformation and the solution of the system of linear ODEs (\ref{nlv01}) only for specific forms of $f(x,y,t)$ and $g(x,y,t)$, which are given below.

\noindent{\bf\underline{Case : 1}}

\noindent Solution for Eq. (\ref{nlv02}) can be deduced by the procedure developed earlier for the choice $f(x,y,t)=(a_1(t)x^p+a_2y^q)^m+b_1(t)$ and $g(x,y,t)=(a_1(t)x^p+a_2y^q)^m+b_2(t)$ with the parametric restriction $\frac{p\beta}{\alpha}=q$.  The solution for this case is the same as the one given by Eq. (\ref{gen-sol1-gen}),
except for the fact $U$ and $V$ are now solutions of the $l^{\mbox{th}}$ order linear ODEs (\ref{nlv01}) which can be given as
\begin{eqnarray}
U(t)=\sum_{i=1}^l I_ie^{m_i t},\quad V(t)=\sum_{i=1}^l \hat{I}_i e^{\hat{m}_i t},\label{nsol1}
\end{eqnarray}
where $I_i$ and $\hat{I}_i$, $i=1,2,\ldots,l$ are $2l$ arbitrary constants while $m_i$ and $\hat{m}_i$ are respectively the $l$ roots of the auxiliary equations
\begin{eqnarray}
m^l+c_{11}m^{l-1}+\ldots+c_{1l}=0\quad \mbox{and}\quad
\hat{m}^l+c_{21}
\hat{m}^{l-1}+\ldots+c_{2l}=0.\label{nsol2}
\end{eqnarray}
\noindent{\bf\underline{Case : 2}}

\noindent For the choice $f(x,y,t)=\left(\sum_{i=1}^N a_i(t)x^{p_i}y^{q_i}\right)^m+b_1(t)$ and $g(x,y,t)=\left(\sum_{i=1}^N a_i(t)x^{p_i}y^{q_i}\right)^m+b_2(t)$ and with the parametric restriction $\frac{p_i\beta}{\alpha}+q_i+1=\frac{k}{m},\,\,i=1,2,\ldots,N$, Eq. (\ref{nlv02}) can be solved.  The solution for this case is also of the same form as given by Eq. (\ref{gen-sol3-gen}), except that now $U$ and $V$ are solutions of the Eqs. (\ref{nlv01}) containing $2l$ arbitrary constants as given in (\ref{nsol1}).

 The results can be straightforwardly extended to the $n$-coupled $l^{\mbox{th}}$ order ODEs.  We do not discuss here separately the three coupled $l^{\mbox{th}}$ order ODEs.
\subsection{Arbitrary $n$-coupled $l^{\mbox{th}}$ order ODEs}
Consider a set of $n$ uncoupled linear $l$th order ODEs of the form
\begin{eqnarray}
\bigg(\frac{d^l}{dt^l}+c_{i1}\frac{d^{(l-1)}}{dt^{(l-1)}}
+\ldots+c_{il-1}\frac{d}{dt}+c_{il}\bigg)U_i(t)=0,
\label {nnv04}
\end{eqnarray}
where $c_{ij}$'s, $i=1,2,\ldots n$, $j=1,2,\ldots l$, are arbitrary constants.
The nonlocal
transformation $U_i=x_i^{\alpha_i}e^{\int f_i(x_1,x_2,\ldots,x_n,t)}$ connects (\ref{nnv04}) to the nonlinear
ODE of the form
\begin{eqnarray}
\bigg(D_{i}^{(l)}+c_{i1}D_{i}^{(l-1)}+\ldots+c_{i,l-1}D_{i}^{(1)}
+c_{il}\bigg)x_i=0,\quad i=1,2,\ldots,n,
\label {nnnv02}
\end{eqnarray}
where $D_{i}^{(l)} =\left(\alpha_ix_i^{\alpha_i-1}(\frac{d}{dt}+x_if_i(x_1,x_2,\ldots,x_n,t))\right)^l$.

Again we find that the solution of Eq. (\ref{nnv04}) can be obtained for the following two cases.

\noindent{\bf\underline{Case : 1}}

\noindent $f_i=(\sum_k^n a_{k}x_k^{p_k})^m+b_i(t))$ with the parametric restriction
$\frac{p_1\alpha_n}{\alpha_1}=\frac{p_2\alpha_n}{\alpha_2}=...=p_n$ leads us to the solution in a form exactly similar to Eq. (\ref{nv08}) for the arbitrary $n$-coupled third order ODEs except that $U_i$'s $i=1,2,\ldots,n$, are now solutions of the uncoupled system of $l^{\mbox{th}}$ order linear ODEs (\ref {nnv04}) as
\begin{eqnarray}
U_i(t)=\sum_{j=1}^l I_{ij} e^{m_{ij}t}\label{nlsol}
\end{eqnarray}
where $I_{ij}$, $i=1,2,\ldots,n$ and $j=1,2,\ldots,l$ are arbitrary constants and $m_{ij}$ are solutions of the auxiliary equations 
\begin{eqnarray}
m_{i}^l+c_{i1}m_i^{l-1}+\ldots+c_{il}=0,\quad i=1,2,\ldots,n.
\end{eqnarray}
We note here that for the special choice of parameters 
$c_{i1}=c_{i2}=\ldots=c_{il}=0$, $\alpha_i=1$ and $p_k=m=1$, $k=1,2,\ldots,n$, Eq. (\ref{nnnv02}) reduces to the $n$-dimensional coupled Ricatti chain whose explicit form can be given as
\begin{eqnarray}
\left(\frac{d}{dt}+\sum_{k=1}^na_kx_k\right)^{l}x_i=0,\quad i=1,2,\ldots,n.  \label{nricatti}
\end{eqnarray}
For the choice $n=2$ the above equation reduces to the two dimensional Ricatti chain given in Eqs. (\ref{chaineq1})-(\ref{chazy}).

Similarly for the parametric choice $c_{i1}=c_{i2}=\ldots=c_{il}=0$, $p_k=2$, $k=1,2,\ldots,n$, and $m=\alpha_i=1$, Eq. (\ref{nnnv02}) reduces to the coupled $n$-dimensional Abel chain whose explicit form can be given as
\begin{eqnarray}
\left(\frac{d}{dt}+\sum_{k=1}^na_kx_k^2\right)^{l}x_i=0,\quad i=1,2,\ldots,n.  \label{nabel}
\end{eqnarray}
Note that the above equation reduces to the two dimensional Abel chain given in Eqs. (\ref{abelchain1})-(\ref{abelchain4}) for the choice $n=2$.

\noindent{\bf\underline{Case : 2}} $f_i=(\sum_{j=1}^Na_j(t)\prod_{k=1}^nx_{k}^{p_{jk}})^m+b_i(t)$ with the parametric restrictions, $\displaystyle\sum_{i=1}^n \frac{p_{ij}\alpha_n}{\alpha_i}+1=\frac{\kappa}{m},\,\,j=1,2,\ldots,N$, gives us again the solutions $x_i(t)$ in exactly
the same form as given in Eq. (\ref{nsecondsol}) for the arbitrary $n$-coupled third order equations except now the set $U_i$, $i=1,2,3,\ldots,n$ in (\ref{nsecondsol}) is the solution of the uncoupled linear system of ODEs (\ref{nnv04}) as given in (\ref{nlsol}).  

In addition to the above $n$-dimensional Ricatti and Abel chains (vide (\ref{nricatti}) and (\ref{nabel})) a new $n$-dimensional integrable chain can be identified which can obtained from Eq. (\ref{nnnv02}) with the parametric choice $c_{i1}=c_{i2}=\ldots=c_{il}=0$, $\alpha_i=1$ and it reads
\begin{eqnarray}
\left(\frac{d}{dt}+(\sum_{j=1}^Na_j(t)\prod_{k=1}^nx_{k}^{p_{jk}})^m\right)^{l}x_i=0,\quad i=1,2,\ldots,n,
\end{eqnarray}
and $\sum_{i=1}^n p_{ij}+1=\frac{\kappa}{m},\,\,j=1,2,\ldots,N$.  By choosing $N=1,\,p_{jk}=1$ and $n=1$, one can obtain Eqs. (\ref{newchain1})-(\ref{newchain4}).
\section{Inhomogeneous linear ODEs and associated coupled nonlinear ODEs}
\label{inhomogeneous}
In our studies, on connection between uncoupled linear ODEs and coupled nonlinear ODEs through nonlocal transformations, we have assumed all along that the set of uncoupled linear ODEs are homogeneous with constant coefficients.  However, this is not mandatory for our procedure to hold good.  For example, we can choose a system of $l^{\mbox{th}}$ order uncoupled linear ODEs of the form
\begin{eqnarray}
\hspace{-1cm}\frac{d^lU_i}{dt^l}+c_{1i}(t)\frac{d^{l-1}U_i}{dt^{l-1}}+\ldots+c_{l-1,i}(t)\frac{dU_i}{dt}
+c_{l,i}(t)U_i=F_i(t),\,\,i=1,2,\ldots,n,\label{inhomlin}
\end{eqnarray}
where $c_{ji}(t),\,\,j=1,2,\ldots l$,\,$i=1,2,\ldots,n$ and $F_i(t)$ are functions of `$t$'.  It is well known that the general solution of (\ref{inhomlin}) can be written as
\begin{eqnarray}
U_i(t)=\alpha_{1i}(t)U_{1i}(t)+\alpha_{2i}(t)U_{2i}(t)+\ldots+\alpha_{li}(t)U_{li}(t),
\end{eqnarray}
where $U_{ji}(t)$'s, $j=1,2,\ldots l$ for a given $i\,(=1,2,\ldots n)$ are $l$ linearly independent solutions of the homogeneous part of Eq. (\ref{inhomlin}) and the coefficient functions $\alpha_{ji}(t)$'s can be given by quadratures which involve $l$ arbitrary constants, for a fixed $i$.  So as long as the general solution of (\ref{inhomlin}) is known, then the entire procedure developed in Secs. \ref{section2}-\ref{section7} completely goes through.  The corresponding coupled nonlinear ODEs will have coefficients involving $c_{ji}(t)$'s which will be now functions of $t$.
\section{Final Comments}
In this paper, we have developed a novel method of identifying two classes of integrable coupled nonlinear ODEs of any order from linear uncoupled ODEs of the same order by introducing suitable nonlocal transformations in the latter.  We found that the problem of solving these classes of coupled nonlinear ODEs of any order, effectively reduces to solving a single first order nonlinear ODE and we have deduced the general solution for the parametric choice for which this first order nonlinear ODE reduces to Bernoulli equation. For suitable choice of parameters we find that the coupled nonlinear ODEs can also exhibit isochronous behaviour.

In this paper we have focussed our attention only on identifying coupled integrable equations through specific type of nonlocal transformations connecting dependent variables only.  There are several possible generalizations : (i) We have considered only two specific integrable choices of Eq. (\ref{generalriccati}).  One can look for other possible integrable cases too. (ii) The functions $f$, $g$ in the transformations can be functions of $\dot{x}$, $\dot{y}$ also. (iii) The independent variable can also be transformed and (iv) One may even consider transformations connecting uncoupled integrable nonlinear ODEs and coupled nonlinear ODEs.  All these possibilities can lead to several new classes of integrable coupled nonlinear ODEs.  Some of these problems are being pursued at present.
\label{section8}

\section*{Acknowledgments}
The work forms a part of a research project of MS and an IRHPA
project of ML sponsored by the Department of Science \& Technology
(DST), Government of India.  ML is also supported by an Indian National Science Academy Senior Scientist award.

\begin{tabular}{p{.15cm}p{14cm}}
\footnotesize$^{1}$ &
N. Euler and P.G.L Leach \emph{Theor. Math. Phys.} {\bf 159}, 474 (2009)\\
\footnotesize$^{2}$ &
Jos\`e F Cari\~nena, Partha Guha and Manuel F Ra\~nada  \emph{Nonlinearity} {\bf 22}, 2953 (2009)\\
\footnotesize$^{3}$&
V.K. Chandrasekar, M. Senthilvelan, Anjan Kundu and M. Lakshmanan
\emph{J. Phys. A: Math. Gen.} {\bf 39}, 9743; \emph{J. Phys. A: Math. Gen.} {\bf 39}, 10945 (2006)\\
\footnotesize$^{4}$&
M. Euler, N. Euler and P.G.L Leach  \emph{J. Nonlinear Math. Phys.} {\bf 14}, 290 (2007)\\
\footnotesize$^{5}$&
S. Moyo and P.G.L Leach \emph{J. Math. Anal. Appl.} {\bf 252}, 840 (2000)\\
\footnotesize$^{6}$&
J.S.R Chisholm and A.K Common \emph{J. Phys. A: Math. Gen.} {\bf 20} 5459 (1987)\\
\footnotesize$^{7}$&
P.G.L Leach \emph{J. Math. Phys.} {\bf 26}, 2510 (1985)\\
\footnotesize$^{8}$&
S. Chandrasekhar \emph{An Introduction to the Study of Stellar Structure} (Dover, New York, 1957)\\
\footnotesize$^{9}$&
J.M Dixon and J.A Tuszynski \emph{ Phys. Rev. A} {\bf41}, 4166 (1990)\\
\footnotesize$^{10}$&
V.K. Chandrasekar, M. Senthilvelan and M. Lakshmanan  \emph{J. Phys. A: Math. Theor.} {\bf 40}, 4717 (2007)\\
\footnotesize$^{11}$&
F. Calogero  \emph{Isochronous Systems} (Oxford University Press,Oxford, 2008)\\
\footnotesize$^{12}$&
R. Gladwin Pradeep, V.K. Chandrasekar, M. Senthilvelan and M. Lakshmanan  \emph{J. Phys. A : Math. Theor.}
{\bf 42}, 135206 (2009)\\
\footnotesize$^{13}$&
V.K. Chandrasekar, M. Senthilvelan and M. Lakshmanan  \emph{Phys. Rev.}
{\bf E72}, 066203 (2005)\\
\footnotesize$^{14}$&
R.R. Vance and E.A. Coddington \emph{J. Math. Biol.} {\bf 27}, 491 (1989)\\
\footnotesize$^{15}$&
Z. Amine and R. Ortega \emph{J. Math. Anal. Appl.} {\bf 185}, 477 (1994)\\
\footnotesize$^{16}$&
Teng Zhidong and Yu Yuanhong  \emph{Acta Mathematicae Applicatae Sinica} {\bf 15}, 401 (1999)\\
\footnotesize$^{17}$&
R. Gladwin Pradeep, V. K. Chandrasekar, M. Senthilvelan and M. Lakshmanan, \emph{A nonlocal connection between certain linear and nonlinear ordinary differential equations : Extension to coupled equations}, arXiv 1008.3232 (2010) \\
\footnotesize$^{18}$&
C.M. Cosgrove \emph{Stud. App. Math.} {\bf104}, 1; {\bf104}, 171 (2000)\\
\footnotesize$^{19}$&
J. Chazy \emph{Acta Math.} {\bf34}, 317 (1911)\\
\footnotesize$^{20}$&
R. Halburd \emph{Nonlinearity} {\bf12}, 931 (1999)\\
\footnotesize$^{21}$&
U. Mugan and F. Jrad \emph{J. Nonlinear Math. Phys.} {\bf9}, 282 (2002)\\
\footnotesize$^{22}$&
N. Euler and M. Euler  \emph{J. Nonlinear Math. Phys.}
{\bf11}, 399 (2004)\\
\footnotesize$^{23}$&
V.K. Chandrasekar, M. Senthilvelan and M. Lakshmanan
\emph{Proc. R. Soc. London A}  {\bf 462}, 1831 (2006)

\end{tabular}
\end{document}